\documentclass[twocolumn,showpacs,preprintnumbers,amsmath,amssymb]{revtex4-1}
\usepackage{graphicx}
\usepackage{rotating}
\usepackage{amssymb}
\usepackage{amsfonts}
\usepackage{amsmath}

\begin{document}

\setcounter{MaxMatrixCols}{10}

\newtheorem{theorem}{Theorem}
\newtheorem{acknowledgement}[theorem]{Acknowledgement}
\newtheorem{algorithm}[theorem]{Algorithm}
\newtheorem{axiom}[theorem]{Axiom}
\newtheorem{claim}[theorem]{Claim}
\newtheorem{conclusion}[theorem]{Conclusion}
\newtheorem{condition}[theorem]{Condition}
\newtheorem{conjecture}[theorem]{Conjecture}
\newtheorem{corollary}[theorem]{Corollary}
\newtheorem{criterion}[theorem]{Criterion}
\newtheorem{definition}[theorem]{Definition}
\newtheorem{example}[theorem]{Example}
\newtheorem{exercise}[theorem]{Exercise}
\newtheorem{lemma}[theorem]{Lemma}
\newtheorem{notation}[theorem]{Notation}
\newtheorem{problem}[theorem]{Problem}
\newtheorem{proposition}[theorem]{Proposition}
\newtheorem{remark}[theorem]{Remark}
\newtheorem{solution}[theorem]{Solution}
\newtheorem{summary}[theorem]{Summary}
\newenvironment{proof}[1][Proof]{\noindent\textbf{#1.} }{\ \rule{0.5em}{0.5em}}

\renewcommand{\theequation}{\thesection.\arabic{equation}}

\newcommand{\re}{\mathop{\mathrm{Re}}}

\newcommand{\lb}{\label}
\newcommand{\be}{\begin{equation}}
\newcommand{\ee}{\end{equation}}
\newcommand{\bea}{\begin{eqnarray}}
\newcommand{\eea}{\end{eqnarray}}

\title{Abolishing the maximum tension principle}

\author{Mariusz P. D\c{a}browski}
\email{mpdabfz@wmf.univ.szczecin.pl}
\affiliation{\it Institute of Physics, University of Szczecin, Wielkopolska 15, 70-451 Szczecin, Poland}
\affiliation{\it Copernicus Center for Interdisciplinary Studies,
S{\l }awkowska 17, 31-016 Krak\'ow, Poland}

\author{H. Gohar}
\email{hunzaie@wmf.univ.szczecin.pl}
\affiliation{\it Institute of Physics, University of Szczecin, Wielkopolska 15, 70-451 Szczecin, Poland}

\date{\today}

\input epsf

\begin{abstract}

We find the series of example theories for which the relativistic limit of maximum tension $F_{max} = c^4/4G$ represented by the entropic force can be abolished. Among them the varying constants theories, some generalized entropy models applied both for cosmological and black hole horizons as well as some generalized uncertainty principle models.

\end{abstract}


\maketitle

\section{Introduction}
\label{intro}
\setcounter{equation}{0}

According to an early remark by Gibbons \cite{Gibbons2002} and Schiller \cite{schiller}, due to the phenomenon of gravitational collapse and black hole formation, there exists a maximum force or maximum tension limit $F_{max} = c^4/4G$ in general relativity ($c$ - the velocity of light, $G$ - Newton gravitational constant). The fact is known as ``The Principle of Maximum Tension''. This is unlike in Newton's gravity, where the two point masses may approach each other arbitrarily close and so the force between them may reach infinity. The limit can nicely be derived by the application of the cosmic string deficit angle $\phi = (8\pi G/c^4) F$ not to exceed $2\pi$ \cite{Gibbons2002}. It is interesting that the maximum tension limit holds also in string theory, where the tension $T$ is given by the Regge slope parameter $\alpha'$, i.e. $F_{max} \propto T = 1/2\pi \alpha'$. The limit is slightly modified in the presence of the positive cosmological constant \cite{BG2014}.

In this paper we would like to show that there exist some theories in which the Principle of Maximum Tension does not hold and we are able to recover the infinite Newtonian tension limit again. One group of them will be the varying constants theories and another the entropic force theories and black hole thermodynamics within the framework of generalized entropy models. The generalized uncertainty principle (GUP) framework will also be exemplified. Special comment about the Carroll limit of the Principle of Maximum Tension will be presented, too.

\section{Maximum tension and the entropic force}
\label{Maximum}
\setcounter{equation}{0}

The maximum tension or force between the two bodies in general relativity has been claimed to be \cite{Gibbons2002}
\be
F_{max} = \frac{c^4}{4G} .
\label{maxten}
\ee
It is advisable to note that the factor $c^4/G$ appears in the Einstein field equations and is of the order of $10^{44}$ Newtons. If the field equations are presented in the form
\be
T_{\mu\nu} = \frac{1}{8\pi} \frac{c^4}{G} G_{\mu\nu}~~~,
\label{EFE}
\ee
where $T_{\mu\nu}$ is the stress tensor and $G_{\mu\nu}$ is the (geometrical) Einstein tensor, then we can consider their analogy with the elastic force equation:
\be
F = k x~~,
\label{elastic}
\ee
where $k$ is an elastic constant, and $x$ is the displacement. In this analogy, we can think of gravitational waves being some perturbations of spacetime and the ratio $c^4/G$ which appears in (\ref{maxten}) plays the role of an elastic constant. Its large value means that the spacetime is extremely rigid or, in other words, it is extremely difficult to make it vibrate \cite{will}.

We make an observation that similar ratio $c^4/G$ appears in the expression for the entropic force within the framework of
entropic cosmology \cite{Easson}. In order to calculate this force one has to apply the Hawking temperature \cite{Hawking}
\be
T=\frac{\gamma \hslash c}{2\pi k_{B}r_{h}}~~,
\label{T}
\ee
and the Bekenstein entropy \cite{Bekenstein}
\be
\text{ \ }S=\frac{k_{B} c^3 A}{4\hslash G} = \frac{\pi k_{B}c^3}{G\hslash } r_h^2 ,
\label{S}
\ee
on the horizon at $r = r_h$, where $A=4\pi r_{h}^{2}$ is the horizon area, $\hslash $ and $k_{B}$ are
Planck constant and Boltzmann constant, respectively, $\gamma $ is an arbitrary and non-negative parameter of the order of unity $O(1)$.

The entropic force is defined as \cite{Easson}
\be
F_r = - T \frac{dS}{dr_h}
\label{EF1}
\ee
and by the application of (\ref{T}) and (\ref{S}), one gets
\be
F_r = - \gamma \frac{c^4}{G} ,
\label{EF2}
\ee
where the minus sign means that the force points in the direction of increasing entropy. It emerges that up to a numerical factor $\gamma/4$ and the sign, this is the maximum force limit in general relativity. The entropic force is supposed to be responsible for the current acceleration as well as for an early exponential expansion of the universe. There is an extra entropic force term into the Friedmann equation and the acceleration equation which are obtained from Einstein field equations (\ref{EFE}).

\section{Abolishing maximum tension}
\label{tension}
\setcounter{equation}{0}

\subsection{Varying constants}

One way to release the maximum tension limit is when we admit the speed of light $c$ and the gravitational constant $G$ in the formula (\ref{maxten}) to vary. There is a Newtonian mechanics limit of (\ref{maxten}) $c \to \infty$, $G \to 0$ which is one of the corners of the Bronshtein-Zelmanov-Okun cube \cite{okun} and it of course recovers infinite tension $F_{max} \to \infty$. Another interesting limit is the Carroll limit $c \to 0$ or $G \to \infty$ which gives $F_{max} \to 0$ \cite{Leblond,Teitelboim}. The $c \to 0$ limit is also predicted within the framework of loop quantum cosmology (LQC) and one can easily understand why it is called the anti-Newtonian limit \cite{mielczarek}, while $G \to \infty$ is just the strong coupling limit of gravity \cite{strongG}. Bearing in mind gravitational wave analogy  (\ref{elastic}), the elastic constant is zero and no gravitational wave can propagate or in other words, the spacetime is infinitely rigid.

Now consider the theory which allows that $c$ and $G$ vary in time in (\ref{EFE}), (\ref{T}), and (\ref{S}). In cosmology, the horizon also depends on time
$r=r(t)$ so that we can write
\be
dS/dr_{h}=\dot{S}/\dot{r} .
\ee
In the varying constants theories \cite{varcon} the entropic force is given by \cite{paper1}
\begin{equation}
F=-T\frac{dS}{dr_{h}}=-\frac{\gamma c^{4}(t)}{2G(t)}\left[ \frac{3\frac{\dot{%
c}(t)}{c(t)}-\frac{\dot{G}(t)}{G(t)}+2\frac{\dot{r}_{h}}{r_{h}}}{\frac{\dot{r_h}}{r_h}}\right] ,
\label{MF1}
\end{equation}%
where we have applied the Hubble horizon
\be
r_h = \frac{c}{H}
\label{rh}
\ee
and get
\begin{equation}
F=-\frac{\gamma c^{4}(t)}{2G(t)}\left[ \frac{5\frac{\dot{c}(t)}{c(t)}-\frac{%
\dot{G}(t)}{G(t)}-2\frac{\dot{H}}{H}}{\frac{\dot{c}(t)}{c(t)}-\frac{\dot{H}}{%
H}}\right] .
\label{MF2}
\end{equation}%
This reduces to (\ref{EF2}) for $\dot{c}=\dot{G}=0$.

The following conclusions are in order. Namely, if the fundamental constants $c$ and $G$ are really constant, then
the maximum force (\ref{MF2}) reduces to a constant value given by (\ref{EF2}). However, the variability of $c$ and $G$ modifies this claim in a way that the maximum force also varies in time. In particular, it seems to be infinite for a constant horizon value $\dot{r}_h = 0$ which corresponds to a model with $c(t) \propto H(t)$. The entropic force can also become infinite, if the derivatives of $c$ and $G$ are infinite.

\subsection{Modified entropy models}

Another way to abolish the Principle of Maximum Tension even without varying $c$ and $G$ is when one changes the definition of entropy (see the appendix in Ref. \cite{komatsu}). The point is that the Bekenstein entropy defined by (\ref{S}) is the area entropy which means that it is proportional to $r_h^2$ so that taking the derivative of (\ref{EF1}), and multiplying it by Hawking temperature (\ref{T}) gives a constant (and a finite) value of the entropic force (\ref{EF2}).

However, if one applies the volume entropy \cite{komatsu}, then one has the entropy which is proportional to $r_h^3$. An example is a nonadditive entropy \cite{Tsallisnonadd} or nonextensive Tsallis entropy \cite{Tsallisnonext} which generalizes (\ref{S}) to
\be
S_3 \propto r_h^3~~.
\ee
Following Ref.\cite{komatsu}, the correct expression for this volume entropy is
\be
S_3 = \zeta \frac{\pi k_B c^3}{\hslash G} r_h^3,
\ee
which as applied to the entropic force definition (\ref{EF1}) together with the Hawking temperature (\ref{T}) gives
\be
F_{r3} = - T \frac{dS_3}{dr_h} = - \frac{3}{2} \gamma \zeta \frac{c^4}{G} r_h ,
\label{Fr3}
\ee
where $\zeta$ is a dimensional constant. Since $r_h = r_h(t)$ according to (\ref{rh}), then the maximum force may reach infinity again, when the horizon size becomes infinitely large. Of course the same happens, if one of the conditions $c \to \infty$ or $G \to 0$ holds.

Another example is the quartic entropy defined as \cite{komatsu}
\be
S_4 = \xi \frac{\pi k_B c^3}{\hslash G} r_h^4 ,
\ee
where $\xi$ is a dimensional constant. It gives an entropic force in the form
\be
F_{r4} = - T \frac{dS_4}{dr_h} = - 2 \gamma \xi \frac{c^4}{G} r_h^2 .
\label{Fr4}
\ee
Here again $r_h = r_h(t)$ according to (\ref{rh}), and so the maximum force may reach infinity when the horizon size becomes infinitely large and this happens much faster than for the volume entropy entropic force (\ref{Fr3}).

For any generalized entropy which is proportional to $r_h^D$, with $D$ being an appropriate volume dimension, we have
\be
S_D \propto r_h^D , \hspace{0.3cm} F_{rD} \propto r_h^{D-2} .
\ee
For varying constants theories, a common generalization of the formula (\ref{MF2}) and the formulas ({\ref{Fr3}) and (\ref{Fr4}) reads as
\be
F_{rD} = -\frac{\gamma c^{4}(t)}{2G(t)}\left[ \frac{(3+D)\frac{\dot{c}(t)}{c(t)}-\frac{\dot{G}(t)}{G(t)}-D\frac{\dot{H}}{H}}{\frac{\dot{c}(t)}{c(t)}-\frac{\dot{H}}{
H}}\right] \left(\frac{c}{H} \right)^{D-2} .
\label{MFD}
\ee

\subsection{Black holes}

Similar considerations about the entropic force can be performed for black holes whose Hawking temperature and Bekenstein entropy are given by
\bea
T=\frac{\hslash \kappa}{2\pi k_B c},
\label{TBH}\\
S_{bh}=\frac{\pi k_B c^3}{G\hslash} r_+^2,
\label{SBH}
\eea
where $\kappa$ and $r_+$ are the surface gravity and the event horizon of a black hole, respectively. In this way, the volume entropy (Tsallis entropy) and quartic entropy for black holes can be written as \cite{Tsallisnonadd,Tsallisnonext}
\bea
S_3=\lambda \frac{\pi k_B c^3 }{4G\hslash} r_+^3,
\label{S3BH}\\
S_4=\beta \frac{\pi k_B c^3 }{4G\hslash} r_+^4,
\label{S4BH}
\eea
where $\lambda$ and $\beta$ are some dimensional constants.

Since we have defined $r_+$ and as a general event horizon, then we start our discussion with charged Reissner-Nordstr\"om black holes for which
the surface gravity is given by \cite{Visser}
\be
\kappa=\frac{c^2}{r_+^2} \sqrt{\frac{G^2M^2}{c^4}-\frac{GQ^2}{4\pi \varepsilon_0 c^4}},
\label{surgrav}
\ee
and the event horizon by
\be
r_+=\frac{GM}{c^2}+\sqrt{\frac{G^2M^2}{c^4}-\frac{GQ^2}{4\pi \varepsilon_0 c^4}},
\label{r+}
\ee
where $M$ is the mass, $Q$ is the charge, $\varepsilon_0$ is the permittivity of space, and we consider a non-extremal case for which
\begin{equation}
M^2>\frac{Q^2}{4\pi \varepsilon_0 G} .
\end{equation}
Using (\ref{TBH}), (\ref{SBH}), (\ref{S3BH}, and (\ref{S4BH}) we can calculate the entropic force (with constant $c$ and $G$) as follows
\be
F_r=-T\frac{dS_{bh}}{dr_+}=- \frac{c^4}{G} \frac{1}{r_+} \sqrt{\frac{G^2M^2}{c^4}-\frac{GQ^2}{4\pi \varepsilon_0 c^4}},
\ee
and
\be
F_{r3}=-T\frac{dS_{3}}{dr_+}=-\frac{3\lambda c^4}{8G}\left(\sqrt{\frac{G^2M^2}{c^4}-\frac{GQ^2}{4\pi \varepsilon_0 c^4}}\right),
\ee
and
\be
F_{r4}=-T\frac{dS_{4}}{dr_+}=-\frac{\beta c^4}{2G}\left(\sqrt{\frac{G^2M^2}{c^4}-\frac{GQ^2}{4\pi \varepsilon_0 c^4}}\right)r_+.
\ee
In the limit $Q \to 0$ the above formulas reduce to the Schwarzschild black hole case for which the surface gravity is $\kappa$=${c^4}/{4GM}$ and the event horizon is equal to the Schwarzschild radius $r_+ = r_s = {2GM}/{c^2}$. In such a case the entropic forces read as
\bea
F_r &=& -T\frac{dS_{bh}}{dr_s} = - \frac{c^4}{2G},\\
F_{r3}& = & -T\frac{dS_{3}}{dr_s} = - \lambda \frac{3 c^4}{16G} r_s ,\\
F_{r4}& = & -T\frac{dS_{4}}{dr_s} =-\beta \frac{c^4}{4G} r_s^2 .
\eea
It is worth noticing that the maximum force ratio $c^4/G$ is already present in the definition of surface gravity. From above equations, we conclude that the entropic force diverges to infinity when the mass of a black hole (proportional to the horizon radius $r_s \propto M$) tends to infinity for the case of the volume entropy and the quartic entropy which contradicts the Principle of Maximum Tension. However, for Bekenstein entropy, the principle holds. For charged Reissner-Nordstr\"om black holes, the entropic force for the Bekenstein entropy tends to $c^4/2G$ while M goes to infinity, so the Principle of Maximum Tension holds as well. However, for the case of the volume and the quartic entropies, the entropic force diverges to infinity when mass tends to infinity, which is again an example of the Maximum Tension Principle violation.

We have not studied Kerr black holes but we expect that the Maximum Tension Principle may also be abolished for them. However, it is not clear if in the approach of Ref. \cite{kerr}, where the authors introduce the effective spring constant for Kerr black holes (related to the angular momentum) which influence the Hawking temperature and Bekenstein entropy, the principle can be avoided, too. One of the reasons is that in the extremal limit, the Hooke's law provides the force which is consistent with the Maximum Tension Principle.

\subsection{Generalized uncertainty principle}

The generalized uncertainty principle (GUP) modifies the Heisenberg principle at the Planck energies into \cite{GUP}
\be
\Delta x \Delta p = \frac{\hslash}{2} \left[ 1 + \alpha^2 (\Delta p )^2 \right],
\ee
where $x$ is the position and $p$ the momentum, and
\be
\alpha = \alpha_0 \frac{l_{pl}}{\hslash}
\ee
($\alpha_0$ is a dimensionless constant). GUP corrects the Bekenstein entropy (\ref{S}) and the Hawking temperature (\ref{T}) of black holes into ($C=$ const.)
\begin{equation}
S_{GUP}= S+\frac{\alpha^2 \pi}{4}\ln{S}-\frac{(\alpha^2\pi)^2}{8}\frac{1}{S}+...+ C,
\label{SGUP}
\end{equation}
and
\begin{equation}
T_{GUP}=T-\frac{\alpha^2\pi}{2}T^2-4(\alpha^2\pi)^4 T^4.
\label{TGUP}
\end{equation}
The above GUP corrected entropy and temperature can be derived by using the quadratic form of GUP. One can also use the linear GUP and modified dispersion relations for possible other modifications of the Bekenstein entropy and the Hawking temperature but we will not be investigating such a case here. By using the above definitions, we can write the GUP corrected entropic force (\ref{EF1}) as
\begin{equation}
F_{rGUP}=-T_{GUP}\frac{dS_{GUP}}{dr}
\end{equation}
or
\begin{equation}
F_{rGUP}=-[F_{r}+\frac{\alpha^2\pi}{4}\frac{F_{r}}{S}-\frac{\alpha^2\pi}{2}T F_{r}+....] .
\label{FrGUP}
\end{equation}
From (\ref{FrGUP}) one can conclude that the GUP force can be influenced by the Hawking temperature (\ref{T}) and the Bekenstein entropy (\ref{S}) and possibly through their dependence on the running fundamental constants $c$ and $G$ they may cause it to diverge then abolishing the Principle of Maximum Tension in this GUP case.

\section{Summary}
\label{summary}
\setcounter{equation}{0}

In this paper we have studied the Principle of Maximum Tension (which says that the force between the two massive bodies in Einstein relativity cannot reach infinity as in the Newton's theory) in the context of different theories of gravity. Firstly, we noticed that the entropic force applied recently to cosmology as a cause of global acceleration up to a numerical factor is just (up to a factor) the maximum force between the two relativistic bodies surrounded by their horizons. We have further explored the issue of abolishing the Principle of Maximum Tension in a couple of physical cases. It has emerged that it has been possible to avoid the principle if one applies the idea that physical constants which enter the expression for the maximum force - the speed of light $c$ and the gravitational constant $G$ - are supposed to vary. This can be concluded from a generalized entropic force formula (\ref{MF2}) or (\ref{MFD}). It is also possible to abolish the principle if one applies different definitions of entropy than the Bekenstein area entropy which is quadratic in the horizon radius and when multiplied by the Hawking temperature gives a finite value of the entropic force (\ref{EF2}). This is not the case if one applies the volume entropy which is cubic in the horizon radius or the quartic entropy which is quartic in the horizon radius. In both cases the force obtained grows as the horizon radius grows and the force may eventually reach infinity. Similarly, the volume and the quartic entropies, as applied to black holes, give the entropic force which may diverge to infinity when the mass of a black hole (both Schwarzschild and charged Reissner-Nordstr\"om) goes up to infinity. In the last of our cases we have considered the generalized uncertainty principle (GUP) we have noticed that the entropic force may reach infinity when the standard Hawking temperature and Bekenstein entropy are influenced in some special way, for example by the variation of the physical constants.

\section{Acknowledgements}

This project was financed by the Polish National Science Center Grant DEC-2012/06/A/ST2/00395.



\end{document}